%% file: main.tex
\documentclass[fleqn,usenatbib]{mnras}
\usepackage{newtxtext,newtxmath}
\usepackage[T1]{fontenc}
\usepackage{caption}
\usepackage{subcaption}
\usepackage[inline]{enumitem}
\usepackage{sistyle}
\usepackage[dvipsnames]{xcolor}
\usepackage{orcidlink}
\usepackage{ulem}
\SIthousandsep{,}
\DeclareRobustCommand{\VAN}[3]{#2}
\let\VANthebibliography\thebibliography
\def\thebibliography{\DeclareRobustCommand{\VAN}[3]{##3}\VANthebibliography}

\usepackage{subfiles}
\usepackage{standalone}

\usepackage{graphicx}	
\usepackage{amsmath}	

\title[CSST Strong Lens Population]{CSST Strong Lensing Preparation: Forecasting the galaxy-galaxy strong lensing population for the China Space Station Telescope}
\author[Cao et al.]{Xiaoyue Cao\orcidlink{0000-0003-4988-9296}$^{1,2}$\thanks{xycao@nao.cas.cn}, 
Ran Li\orcidlink{0000-0003-3899-0612}$^{1,2}$\thanks{ranl@bao.ac.cn},
Nan Li\orcidlink{0000-0001-6800-7389}$^{2,1}$,
Rui Li\orcidlink{0000-0002-3490-4089}$^{1,2}$,
Yun Chen\orcidlink{0000-0001-8919-7409}$^{2,1}$,
Keyi Ding$^{1,2}$,
Huanyuan Shan$^{3,1}$,
\newauthor
Hu Zhan\orcidlink{0000-0003-1718-6481}$^{2,4}$,
Xin Zhang$^2$,
Wei Du$^5$,
\& Shuo Cao$^6$
\\
$^{1}$School of Astronomy and Space Science, University of Chinese Academy of Sciences, Beijing 100049, China\\
$^{2}$National Astronomical Observatories, Chinese Academy of Sciences, 20A Datun Road, Chaoyang District, Beijing 100012, China\\
$^{3}$ Shanghai Astronomical Observatory (SHAO), Nandan Road 80, Shanghai 200030, China \\
$^{4}$ Kavli Institute for Astronomy and Astrophysics, Peking University, Beijing, 100871, China \\
$^{5}$ Shanghai Key Lab for Astrophysics, Shanghai Normal University, Shanghai, 200234, China \\
$^{6}$ Department of Astronomy, Beijing Normal University, 100875, Beijing, China
}

\date{Accepted XXX. Received YYY; in original form ZZZ}

\pubyear{2015}

\begin{document}
\label{firstpage}
\pagerange{\pageref{firstpage}--\pageref{lastpage}}
\maketitle

\begin{abstract}
Galaxy-galaxy strong gravitational lensing (GGSL) is a powerful probe for the formation and evolution of galaxies and cosmology, while the sample size of GGSLs leads to considerable uncertainties and potential bias. The China Space Station Telescope (CSST, to be launched in late 2026) will conduct observations across 17,500 square degrees of the sky, capturing images in the $ugriz$ bands with a spatial resolution comparable to that of the Hubble Space Telescope. We ran a set of Monte Carlo simulations to predict that the CSST's wide-field survey will observe $\sim$160,000 galaxy-galaxy strong lenses over its lifespan, increasing the number of existing galaxy-galaxy strong lens samples by three orders of magnitude. This is comparable to the capabilities of the $\it Euclid$ telescope but with the added benefit of additional color information. Specifically, the CSST can detect strong lenses with Einstein radii about $0.64\pm0.42 \arcsec$, corresponding to the velocity dispersions of $217.19 \pm 50.55 \, \text{km/s}$. These lenses exhibit a median magnification of $\sim$5. The apparent magnitude of the unlensed sources in the g-band is $25.87 \pm 1.19$. The signal-to-noise ratio of the lensed images covers a range of $\sim 20$ to $\sim 1000$, allowing us to determine the Einstein radius with an accuracy ranging from $\sim 1 \%$ to $\sim 0.1 \%$, ignoring various modeling systematics. Our estimates indicate that CSST can observe rare systems like double source-plane and spiral galaxy lenses. The above selection functions of the CSST strong lensing observation help optimize the strategy of finding and modeling GGSLs.

\end{abstract}




\subfile{intro.tex}
\subfile{method.tex}
\subfile{result.tex}

\subfile{discuss.tex}

\subfile{summary.tex}

\section*{Acknowledgements}
We thank the referee, Conor O'Riordan, for the helpful suggestions that have improved the paper to its current form. This work is supported by the National Key R\&D Program of China (grant number 2022YFF0503403), the National Natural Science Foundation of China (No. 11988101), the K.C.Wong Education Foundation, and the science research grants from China Manned Space Project with No.CMS-CSST-2021-B01. XYC acknowledges the support of the National Natural Science Foundation of China (No. 12303006) and the valuable suggestions offered by Yiping Shu. RL is also supported by the National Natural Science Foundation of China (Nos 11773032, 12022306). W.D. acknowledges the support from the National Natural Science Foundation of China (NSFC) under grant 11890691. We express our gratitude to ChatGPT, an AI model developed by OpenAI, for its assistance in polishing the English of this paper. This research was initiated under the support of the ISSI/ISSI-BJ International Team Programs.

\section*{Data Availability}
The code and data product that supports this work is publicly available from \url{https://github.com/caoxiaoyue/sim_csst_lens}.




\bibliographystyle{mnras}
\bibliography{reference} 




\appendix
\subfile{appx_A.tex}

\subfile{appx_B.tex}
\subfile{appx_C.tex}
\subfile{appx_D.tex}
\subfile{appx_E.tex}

\bsp	
\label{lastpage}
\end{document}

%% file: intro.tex
\section{Introduction}
\label{sec:intro}
Galaxy-galaxy strong gravitational lensing is a phenomenon in which the gravitational field of a foreground lens galaxy significantly distorts the light emitted from a background source galaxy, leading to the formation of multiple images or extended arc structures. Lens modeling enables the reconstruction of both the mass distribution of the lens and the intrinsic brightness distribution of the source based on the observed multiple images. Furthermore, the time delay between different lensed images depends on the spacetime background of the universe, allowing us to utilize this information to infer the cosmological parameters. Over the past few decades, strong lensing has emerged as a potent tool for astronomers, serving as a `cosmic telescope' to observe magnified high-redshift sources \citep{Newton11,Shu16,Cornachione18,Blecher19,Ritondale19_src,Rizzo20,Cheng20,Marques20,Yang21, Chak2023}, probing the mass (sub)structure of the lens galaxy \citep{lmass_Treu06,lmass_Koopmans06,lmass_Gavazzi07,lmass_Bolton08,lmass_Vegetti09,lmass_Auger10_lens,lmass_Bolton12,lmass_he18,lmass_Nightingale19,slope_chen19,lmass_he20,lmass_Du20, Nightingale2022}, providing constraints on the nature of dark matter \citep{DM_Mao1998,DM_Vegetti12,DM_Vegetti14,DM_RanLi16,DM_RanLi17,DM_Ritondale19,DM_Gilman19,DM_Gilman20,DM_He20,Enzi20,Bayer2023,Vegetti2023}, and testing the theory of General Relativity on galactic scale \citep{Bolton_ppn, Cao_ppn, Collet_ppn, Yang_ppn}.
It also acts as an alternative cosmography tool, facilitating measurements of the Hubble constant, matter density parameters, and exploration of various dark energy models \citep{Suyu13, slope_chen19, Birrer20, wong20, Millon20, Hui23, Tian2024}.

The number of strong gravitational lens candidates is relatively small, amounting to only a few thousand, of which only several hundred have been spectroscopically confirmed. Consequently, various aspects of strong gravitational lensing research necessitate larger sample sizes to achieve their scientific goals. For instance, a dataset of five thousand GGSLs would alleviate the degeneracy between stellar mass and the inner density slope of dark matter halos, leading to tighter constraints on the galaxy's initial mass function \citep{Collett23}. By selecting hundreds of high-quality GGSLs capable of detecting dark matter subhalos weighing approximately $10^7 M_{\odot}$, we can either validate or challenge the cold dark matter model \citep{DM_RanLi16}. Moreover, with around 40 strongly lensed quasar systems that possess high-quality imaging and spectroscopic data, the Hubble constant can be constrained to a precision of $\sim 1\%$ \citep{Shajib18_ifu}, allowing us to decisively determine if a discrepancy truly exists between the early and late cosmographic measurements \citep{akin20, Birrer22_review, Treu22}. While comprehending potential systematics remains crucial for achieving these scientific objectives, future strong lensing studies will greatly benefit from an expanding sample size.

In addition to enlarging the sample size, future surveys can enhance sample completeness, specifically to identify either rare or undetected lenses within the existing sample. These include lower-mass lenses \citep{Shu17}, higher-redshift lenses \citep{Jacobs19}, and additionally, more exotic systems, such as double-source plane lenses \citep{Collett14}, and strongly lensed supernovae \citep{lensed_SNE_1, lensed_SNE_2, lensed_SNE_3}. The statistics of strong lensing will undoubtedly reach unprecedented levels with the enhancement of sample completeness \citep{Sonnenfeld21a,Sonnenfeld21b,Sonnenfeld22c}.

In recent years, researchers have expanded the number of galaxy-galaxy strong lenses to several thousand by mining the massive data from current ground-based image surveys, such as KIDS \citep{kids_survey}, DESI \citep{DESI_image_survey}, and HSC \citep{HSC_survey}, with the assistance of automated image recognition powered by machine learning techniques \citep{kids_lens_1, kids_lens_2, kids_lens_3, desi_lens_1, desi_lens_2, desi_lens_3, hsc_lens_1, hsc_lens_2, hsc_lens_3, hsc_lens_4}. Nonetheless, ground-based telescopes face difficulties in detecting lenses with small Einstein radii (or small masses) due to atmospheric seeing\footnote{For ground-based observations, the Full-Width-Half-Maximum (FWHM) of the Point Spread Function (PSF) is typically larger than $\sim0.7$ arcsec.}, and their image data may not permit high-precision lensing models. The Euclid survey \citep{euclid_survey}, which was launched in 2023, aims to image approximately 15,000 square degrees of the sky using VIS/NISP cameras. It is anticipated to detect around 150,000 galaxy-galaxy strong lenses \citep{Collett15}. The survey's high spatial resolution (PSF FWHM, approximately $0.18 \arcsec$) provides a distinct advantage over ground-based surveys and holds scientific potential for advancing large lens sample research. Because Euclid does not offer color information in the optical band, it becomes challenging to carry out robust photometric redshift measurements to determine the lens's and the source's redshift. Instead, spectroscopy follow-up is necessary (which can be costly for many lenses) to ensure that the ``angular quantities'' measured from the lensing image can be accurately interpreted as physical quantities. Additionally, the absence of color information hinders Stellar Population Synthesis (SPS) analysis, impeding the decomposition of different mass components (stellar and dark matter) in lens galaxies \citep{Auger09}.

The China Space Station Telescope \citep[CSST,][]{csst_zhanhu}, a 2-meter survey telescope, is scheduled to launch in late 2026. It will offer image resolution comparable to the Hubble Space Telescope (PSF FWHM, approximately $0.07\arcsec$) in seven bands, including $u$, $g$, $r$, $i$, $z$, $y$, and $NUV$. Over a 10-year lifespan, CSST will survey approximately 17,500 square degrees of the sky, making it a valuable resource for large lens sample studies. This work aims to comprehensively understand the population characteristics and selection function for strong lensing observations conducted by CSST. To achieve this, a suite of Monte Carlo simulations is employed, considering a range of observational effects caused by the CSST instrument. This study will lay the foundation for future strong lensing research using CSST.

This paper is organized as follows. We present the method to simulate the CSST strong lens sample in Section~\ref{sec:method} and the corresponding results in Section~\ref{sec:results}. We discuss the uncertainties associated with the predicted number of lenses and the future prospects of CSST strong lensing science in Section~\ref{sec:discuss}. A brief summary is given in Section~\ref{sec:summary}. The flat $\Lambda$CDM cosmology is assumed throughout this work, with parameter values of $\Omega_m = 0.3$, $\Omega_\Lambda = 0.7$, and $H_0 = 70\, \mathrm{km}\,\mathrm{s}^{-1}\,\mathrm{Mpc}^{-1}$.


%% file: method.tex
\section{Methodology}
\label{sec:method}
To predict the properties of galaxy-galaxy strong gravitational lenses (GGSLs) observed in a specific image survey, understanding the population characteristics of the foreground and background galaxies is essential. When the background galaxy resides within the ``Einstein light cone'' of the foreground galaxy, a strong lens system is formed, with the background galaxy serving as the source and the foreground galaxy as the deflector. The ``Einstein light cone'' refers to the region behind the foreground galaxy where the source is strongly lensed. Moreover, it is crucial to account for the transfer function \citep{transfer_func_def} of the CSST GGSL observations. This transfer function describes the mapping from all strong lenses distributed across the entire sky (i.e., the ideal GGSLs) to the ones that are detectable in the actual survey. The method taken in this work incorporates the population model of foreground and background galaxies utilized in \cite{Collett15}. This model is justified by its ability to reproduce the observed population of GGSLs from the SL2S survey, as shown in Figure 3 of their paper. A brief review of the population model used in this work is presented in Section~\ref{sec:lens_pop}. Section~\ref{sec:sim_ideal_lens} outlines the methodology for simulating the ideal GGSLs across the entire sky. Section~\ref{sec:transfer_func} introduces the two key factors that significantly influence the transfer function: the instrument qualification of CSST and the choice of the lens finder. The code supporting this work is publicly accessible at \url{https://github.com/caoxiaoyue/sim_csst_lens}.

\subsection{The Lens population}
\label{sec:lens_pop}
Although we adopt the empirical relations from \cite{Collett15} without any modifications \footnote{See also the \texttt{LensPop} code: \url{https://github.com/tcollett/LensPop}.}, we provide a concise summary for the sake of self-consistency. For more comprehensive information, we refer readers to \cite{Collett15} and the references therein.

\subsubsection{Foreground galaxy population}
\label{sec:foreground_pop} 
We assume that the population of foreground galaxies consists solely of Early-Type Galaxies (ETGs), thereby excluding lensing systems in which the deflectors are non-ETGs (such as spiral and irregular galaxies). Given that the lensing cross section of massive ETGs is typically higher than others, we expect that this selection does not substantially underestimate the number of observable lenses \citep{moller07}. The mass distribution of ETGs can be represented by the Singular Isothermal Ellipsoid model (SIE) \citep{Kormann_sie}. The SIE model is characterized by five parameters: the Einstein radius ($\theta_{\rm E}$), the axial ratio ($q_f^m$), the position angle ($\phi_f^m$), and the center ($x_f^m$, $y_f^m$), where the superscript and subscript denote the ``mass'' distribution and the ``foreground'' nature of the galaxy, respectively. The Einstein radius ($\theta_{\rm E}$) is directly related to the velocity dispersion ($\sigma_V$) via\footnote{We assume that the Einstein radius derived from Equation~\eqref{eq:thetaE_vdisp} is applicable to the singular isothermal ellipsoid model, following the intermediate axis convention \citep{Shu2015_intermediate_axis}.},
\begin{equation}
\theta_{\mathrm{E}}^{\mathrm{SIS}}=4 \pi \frac{\sigma_V^2}{c^2} \frac{D_{\mathrm{ls}}}{D_{\mathrm{s}}} .
\label{eq:thetaE_vdisp}
\end{equation}
where $D_{\rm ls}$ and $D_{s}$ represent the angular distance from the lens and the observer to the source, respectively; $c$ is the speed of light. and $\sigma_V$ obeys the following velocity dispersion function \citep{sdss_vdf} 
\begin{equation}
d n=\phi_*\left(\frac{\sigma}{\sigma_*}\right)^\alpha \exp \left[
-\left(\frac{\sigma}{\sigma_*}\right)^\beta\right] \frac{\beta}{\Gamma(\alpha / \beta)} \frac{d \sigma}{\sigma},
\label{eq:vdisp_func}
\end{equation}
where $\phi_*=8.0 \times 10^{-3} h^3 \mathrm{Mpc}^{-3}$, $\sigma_*=161 \mathrm{~km} \mathrm{~s}^{-1}$, $\alpha=2.32$, $\beta=2.67$, are best-fitting values derived from the analysis of local ETG data in SDSS DR5; and $d n$ represents the number density (in terms of the comoving volume) of ETGs per velocity interval. We assume that the velocity dispersion function remains constant with redshift, consistent with the observational findings presented by \cite{vdf1, vdf2, vdf3, vdf4}, and is truncated at $\rm 50 \, \text{km/s}$. The axis-ratio $q_f^m$ can be randomly drawn from the Rayleigh distribution if the $\rm \sigma_V$ is known, i.e.
\begin{equation}
P\left(1-q \mid s=\left(A+B \sigma_V\right)\right)=\frac{1-q}{s^2} \exp \left(\frac{-(1-q)^2}{2 s^2}\right),
\label{eq:foreground_q}
\end{equation}
where the values of $A = 0.38$ and $B = 5.7 \times 10^{-4}$ are obtained through fitting the SDSS data \citep{Collett15}.
Equation~\eqref{eq:foreground_q} indicates that a galaxy with a higher velocity dispersion (mass) tends to have a more circular shape. The translational and rotational symmetry of space enables us to position the foreground galaxy at the center of the image ($x_f^m=0$, $y_f^m=0$) and align the major axis of the SIE with the x-axis of the image ($\phi_f^m=0$).

The light distribution of the foreground galaxy is modeled using the \textit{de Vaucouleurs model}, which incorporates six parameters: the central position ($x_f^l$ and $y_f^l$), the position angle ($\phi_f^l$), the axis ratio ($q_f^l$), the half-light radius ($R_{\rm e}^l$), and the absolute magnitude ($\rm M_f$). The superscript $l$ denotes the ``light'' distribution of the galaxy. The morphological parameters, namely $x_f^l$, $y_f^l$, $\phi_f^l$, $q_f^l$, and $R_{\rm e}^l$, are assumed to be consistent across all bands, except for the absolute magnitude ($\rm M_f^X$), which varies depending on the band denoted by the superscript $X$. The light trace mass assumption is made, leading to $x_f^l \equiv x_f^m = 0$, $y_f^l \equiv y_f^m = 0$, $\phi_f^l \equiv \phi_f^m=0$, and $q_f^l \equiv q_f^m$. The absolute magnitude in the $r$ band ($\rm M_f^r$) and the $R_{\rm e}^l$ in kilo-parsec unit can be derived from $\rm \sigma_V$ using the following equations
\begin{equation}
\begin{aligned}
& V = \log_{10} \sigma_V \\
& M_f^r=\frac{-0.37+\sqrt{0.37^2-(4\times0.006\times(2.97+V))}}{2\times0.006} + \mathcal{N}(0.15/2.4) \\
& R_{\rm e}^l = 2.46- 2.79 \times V + 0.84\times V^2 + \mathcal{N}(0.11)
\end{aligned} 
\label{eq:foreground_Re_M}
\end{equation} 
where $\mathcal{N}(0.11)$ denotes a Gaussian distribution with a standard deviation of 0.11. Equation~\eqref{eq:foreground_Re_M} utilizes the fundamental plane relation as provided by \cite{Hyde09}. Absolute magnitudes of foreground galaxies in bands other than $r$ (representing the color) can be obtained by extrapolating the template spectra of passive galaxies that underwent a single burst of star formation 10 Gyrs ago.

In summary, the population model of foreground galaxies employed in this work makes the assumption that the properties of ETGs, such as the axis ratio, the Einstein radius, the r-band absolute magnitude, and the half-light radius, are exclusively governed by the velocity dispersion. The velocity dispersion is sampled randomly from the distribution described by Equation~\eqref{eq:vdisp_func}. The redshift of foreground galaxies is truncated at 2.5\footnote{Further increasing the redshift truncation value does not significantly alter the main results reported in this work.}, which is sufficiently deep to predict the GGSLs (see Figure~\ref{fig:ideal_len}).

\subsubsection{Background galaxy population}
\label{sec:background_pop}
We assume that the brightness distribution of background galaxies follows an Elliptical Exponential Disk (EED) model\footnote{The exponential disk model is a specific instance of the Sersic model \citep{Sersic1963}, with the index parameter set to 1.} \citep{disk_model}, which is a reasonable assumption given that high redshift sources are typically star-forming galaxies \citep{Newton11}.
The EED model is characterized by six parameters: the central position ($x_s$, $y_s$), the axis ratio ($q_s$), the position angle ($\phi_s$), the half-light radius ($R_{\rm e}^{s}$), and the absolute magnitude at band X ($M_s^X$). The parameters $x_s$, $y_s$, and  $\phi_s$ are assumed to be randomly distributed across the sky. The axis ratio $q_s$ follows a Rayleigh distribution with a scale parameter of $s = 0.3$ and is truncated at 0.2. For a source with a given magnitude at band $r$ ($M_s^r$), we randomly draw the corresponding values of $R_{\rm e}^{s}$ using the empirical relation proposed by \cite{Mosleh12, Huang13}
\begin{equation}
\log _{10}\left(\frac{R_{\rm e}^{s}}{\mathrm{kpc}}\right)= -\frac{M^r_s+18}{4} + \log _{10}\left(\frac{1+z}{1.6}\right)^{-1.2}+N(0.35),
\label{eq:background_Re}
\end{equation}
where $z$ denotes the redshift of the source. Similar to the approach adopted for the foreground galaxies, we assume that the half-light radius remains consistent across all bands while only the magnitude varies. Thus, the distribution of $M_s^X$ represents the final component required to describe the morphology of the background sources. Current observations have not provided sufficient depth to obtain the luminosity function of faint galaxies, and neglecting these faint galaxies in lensing simulations would result in a substantial underestimation of the number of observable GGSLs. Therefore, we use the mock source catalog produced by the cosmological simulation \cite{Connolly10}, which has also been used by \cite{Collett15} to forecast the lensing population for the Euclid survey. This source catalog is generated by ray-tracing through the Millennium $N$-body simulation \citep{springel2005}, with galaxies added using a semi-analytic model developed by \cite{de2006}. The cosmological simulation has been calibrated to reproduce the luminosity and color distributions of low-redshift galaxies, as well as the redshift and number-magnitude distributions derived from deep imaging and spectroscopic surveys. The source catalog covers a 4.5$\times$4.5 square degrees region of the sky, spanning a redshift range of approximately $z=0.2$ to 5. It is complete down to an $i$-band magnitude of 27.5, thus providing sufficient depth for simulating the distribution of GGSLs for the CSST.

\subsection{Simulating the ideal lenses}
\label{sec:sim_ideal_lens}
We assume a uniform distribution of foreground ETGs in comoving space, with a velocity dispersion distribution defined by Equation~\eqref{eq:vdisp_func}. To determine the number of GGSLs in the sky, we need to answer the following question: What is the probability of a given ETG undergoing strong lensing? In other words, what is the likelihood of another background source galaxy coincidentally falling within the ``Einstein light cone'' of the foreground ETG?

Considering the source catalog used in this study, the number density of sources is approximately 0.06 objects per square arcsecond. This implies that for any ETG selected according to the empirical law described in Section~\ref{sec:foreground_pop}, if we place a square box centered on the ETG's position with sides of $\sqrt{1/0.06} \approx 4\arcsec$, the box, on average, contains one source galaxy. Assuming the source is randomly distributed within the box, the conditions for an ETG-source pair to form a strong lensing system can be expressed as follows:
\begin{enumerate*}[label=(\roman*)]
\item $z_s > z_f$ (the source is behind the ETG),\label{item:z_cond}
\item $x_s^2+y_s^2 < \theta_E^2$ (the source galaxy falls inside the ``Einstein light cone'' of the ETG). \label{item:strong_lens_cond}.
\end{enumerate*}

\subsection{The transfer function}
\label{sec:transfer_func} 
The ideal GGSLs defined in Section~\ref{sec:sim_ideal_lens} may not be identifiable in real imaging surveys. The limitations primarily arise from the finite spatial resolution resulting from the granularity of CCD pixels and the blurring effect of the Point Spread Function (PSF). Moreover, the detectability of GGSLs is also influenced by the limited depth of the survey, which is determined by factors such as instrumental zero-point magnitude, exposure time, sky background level, and readout noise. Lastly, the choice of the lens finder can further impact the identification of these lenses. All the aforementioned factors that influence the observability of GGSLs are encompassed within the ``transfer function'' that describes the mapping from the ideal GGSL sample to the observable one.

In this work, we simulate realistic lensing images under the CSST survey using the instrumental properties presented in Table~\ref{table:surveys}. We then utilize the lens finder described in \cite{Collett15} to determine if a lens is detectable. The lens finder we employ adheres to the following criteria:
\begin{enumerate*}[label=(\roman*)]
\item The total magnification $\mu$, defined as the ratio of the total flux between the lensed and unlensed sources, must be greater than 3; \label{item:magnification_cond}
\item The tangential stretching of the source ($\mu R_{\rm e}^{s}$) must be larger than the FWHM of the PSF $s$; \label{item:tangent_shear_cond}
\item The multiple images can be spatially resolved, $\theta_{\mathrm{E}}^2 > (R_{\rm e}^{s})^2 + (s / 2)^2$; \label{item:resolved_cond}
\item The total signal-to-noise ratio (SNR) of the lensed images, denoted as $\mathrm{SNR}_{\text{arc}}$, must be greater than 20. $\mathrm{SNR}_{\text{arc}}$ is calculated by quadrature summation of all pixels in the lensed images with an SNR greater than 1.
\item The Einstein radius is larger than the size of a single pixel, specifically $0.074 \arcsec$.
\end{enumerate*}
The impact of using alternative strong lens finders on predicting the CSST GGSL sample is discussed in Section~\ref{sec:discuss}.

\begin{table*}
 \caption{The key properties of the CSST imaging survey include the survey area ($\Omega$), the filters used for lensing simulation, the Full-Width at Half Maximum (FWHM) of the Point Spread Function (PSF), the CCD gain in electrons per ADU, the median sky brightness in magnitude per square arcseconds, the readout noise, the total exposure time in each filter (calculated by multiplying individual exposure time with the number of exposures), the image pixel scale, the instrumental zero-point, and the $5\sigma$ limiting magnitude for a point-like source (within $R_\mathrm{EE80}$, the radius of 80\% encircled energy). Furthermore, we include the relevant properties of Euclid \citep{euclid_instru_1, euclid_instru_2} for comparison.
\label{table:surveys}} 
\begin{tabular}{ccccc}
\hline
                       & CSST-WF                                        & CSST-DF     & CSST-UDF            & Euclid                                            \\
\hline
$\Omega$ {[}deg$^2${]} & 17500                                          & 400         & 9             & 15000                                          \\
Filters                & \{$g,r,i,z$\}                                    & -           & -             & \{$VIS$\}                                  \\
PSF FWHM {[}arcsec{]}  & \{0.051, 0.064, 0.076, 0.089\}$^{a}$                           & -           & -             & \{0.18\}                            \\
Gain    [$e^-$/ADU]               & 1.5                      & -           & -             & 3.1                         \\
Sky background         & \{22.57, 22.10, 21.87, 21.86\}                         & -           & -             & \{22.35\}                     \\
Readout noise {[}$e^{-1}${]}        & 5                                 & -           & -             & 4.5                                                \\
Exposure time {[}s{]}        & {[}$150\times2$, $150\times2$, $150\times2$, $150\times2${]} &  {[}$250\times8$, $250\times8$, $250\times8${]}   & {[}$250\times60$, $250\times60$, $250\times60${]} & {[}$565\times3${]}$^b$ \\
Pixel size {[}arcsec{]}    & 0.074                                          & -           & -                & 0.1   \\
Zero point             &  \{25.79, 25.60, 25.41, 24.83\}                      & -           & -             & \{24\}                           \\
Limiting magnitude & \{26.58, 26.32, 26.03, 25.46\}       &  \{27.77, 27.50, 27.20, 26.67\} & \{28.89, 28.62, 28.32, 27.78\} & \{27.27\}$^{cd}$ \\
\hline
\end{tabular}
\newline\footnotesize
\flushleft{$^a$ The requirement of CSST PSF is defined as $R_{EE80}<0.15 \arcsec$ at $\lambda = 632.8$ nm. For comparison with Euclid, the FWHM values are determined by fitting a Gaussian model to the simulated PSF image of CSST. Please refer to Appendix~\ref{sec:appx_a} for additional context.}
\flushleft{$^b$ We assume that all data collected by the Euclid survey will have only three exposures, despite the fact that about half of the data is expected to have four exposures. As a result of this assumption, our predictions for the number of strong lenses detected by the Euclid survey may be underestimated.}
\flushleft{$^c$ It is important to note that the reported magnitude corresponds to the $5\sigma$ limiting magnitude for a point source measured within the $R_\mathrm{EE80}$, rather than the one typically defined for sources with an extent of $\sim 0.3 \arcsec$ \citep{euclid_instru_1}. To calculate this limiting magnitude value, we make the assumption that Euclid’s PSF follows a Gaussian form, meaning its $R_\mathrm{EE80}=0.137 \arcsec$. 
}
\flushleft{$^d$ For a more comprehensive discussion on comparing the depth of CSST and Euclid, see Appendix~\ref{sec:appx_c}.}
\end{table*}

%% file: result.tex
\section{Results}
\label{sec:results}
We present the ideal lens sample population in Section~\ref{sec:res_ideal_lens}. Our predictions for the GGSL sample of CSST are shown in Section~\ref{sec:res_csst_lens}. Section~\ref{sec:res_lens_comparison} compares the GGSL sample of CSST with other optical image surveys conducted during the same period, such as the Euclid and LSST.

\begin{figure*}
  \centering
    \includegraphics[width=0.9\textwidth]{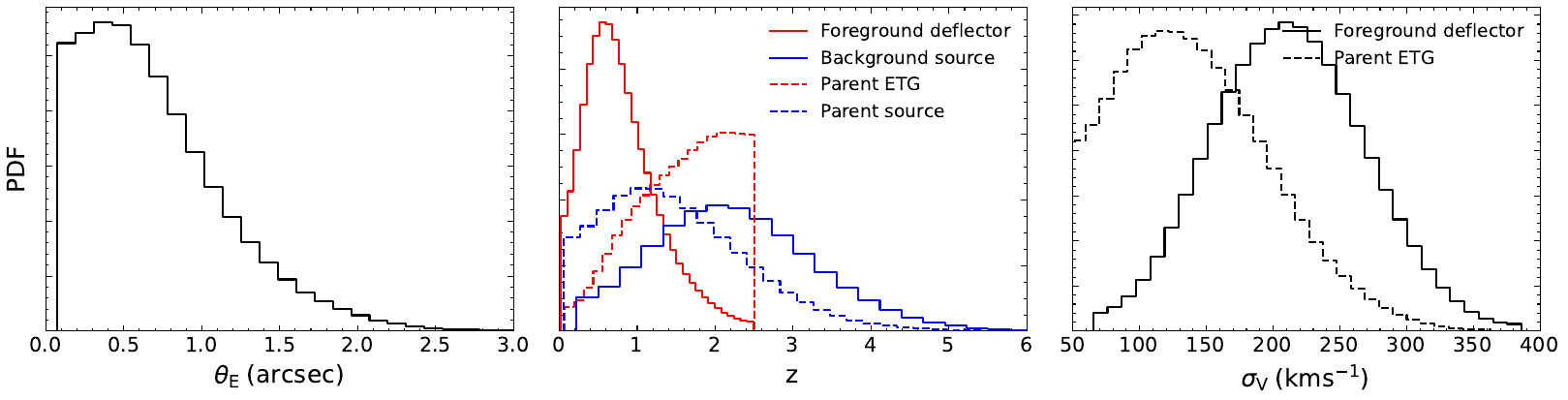}
\caption {Population statistics of the ideal strong lens sample. \textit{Left panel}: The probability density distribution of the lens's Einstein radius. \textit{Middle panel}: The redshift probability density distribution. The red and blue histograms represent the distributions of the foreground lens and the background source, respectively, while the red and blue dashed lines represent the distributions of the parent ETG and the source used to simulate strong lenses, respectively. \textit{Right panel}: The probability density distribution of the ETG’s velocity dispersion.}
\label{fig:ideal_len}
\end{figure*}

\subsection{Ideal lens sample}
\label{sec:res_ideal_lens}
Based on the ETG population described in Section~\ref{sec:foreground_pop}, our simulation suggests that the entire sky harbors approximately 2.1 billion ETGs within the redshift range of 0 to 2.5. Subsequently, with the introduction of the source population discussed in Section~\ref{sec:background_pop}, approximately 12 million of these ETGs act as deflectors, exerting a strong lensing effect on the background source. The rate of strong lensing is estimated to be around 0.6\%. Figure~\ref{fig:ideal_len} depicts the population properties of the ideal lens sample. The left panel displays the probability density distribution of the lensing Einstein radius, peaking at approximately 0.5$\arcsec$, with a tail extending to $3.0 \arcsec$. In the middle panel, the red solid line signifies the redshift probability density distribution of the lens galaxies, whereas the blue solid line represents the distribution of the source galaxies. The lens galaxies display a median redshift of approximately 0.71. The red and blue dashed lines correspond to the redshift distributions of the original ETGs and sources used for lensing simulation, respectively. In this work, we manually truncate the redshift distribution of the original ETGs at 2.5. Considering the scarcity of lensing objects with redshifts near 2.5 in the ideal lens sample (as evident from the red solid line), this choice does not lead to a significant underestimation of the forecasted number of lens samples, at least concerning the source catalog employed in this study. The solid line in the right panel shows the probability density distribution of the velocity dispersion for the lens galaxy, having a median of $210.53 \pm 52.34 \, \text{km/s}$; the dashed line shows the parent galaxy sample. It is worth noting that our Figure~\ref{fig:ideal_len} is in agreement with the results presented in \cite{Collett15}, thereby validating the simulation code utilized in this work.

\subsection{CSST lens sample}
\label{sec:res_csst_lens}
The CSST survey comprises three operational modes: the ``Wide Field'' (WF), the ``Deep Field'' (DF), and the ``Ultra Deep Field'' (UDF). The WF mode covers a sky area of 17,500 square degrees with a $300\,s$ exposure time. In contrast, the DF mode uses a longer exposure time of $2,000\,s$ but covers a smaller survey area of 400 square degrees. Additionally, the UDF mode aims to achieve a similar scientific goal as the well-known Hubble Ultra-Deep Field by targeting a sky area of 9 square degrees with an exposure time of $15,000\,s$. Sections~\ref{sec:WF_mode}, \ref{sec:DF_mode}, and \ref{sec:UDF_mode} present the GGSLs of CSST under the WF, DF, and UDF surveys, respectively.

\subsubsection{Wide-field lens sample}
\label{sec:WF_mode}
\begin{figure*}
  \centering
  \begin{subfigure}[t]{\textwidth}
    \centering
    \includegraphics[width=0.9\textwidth]{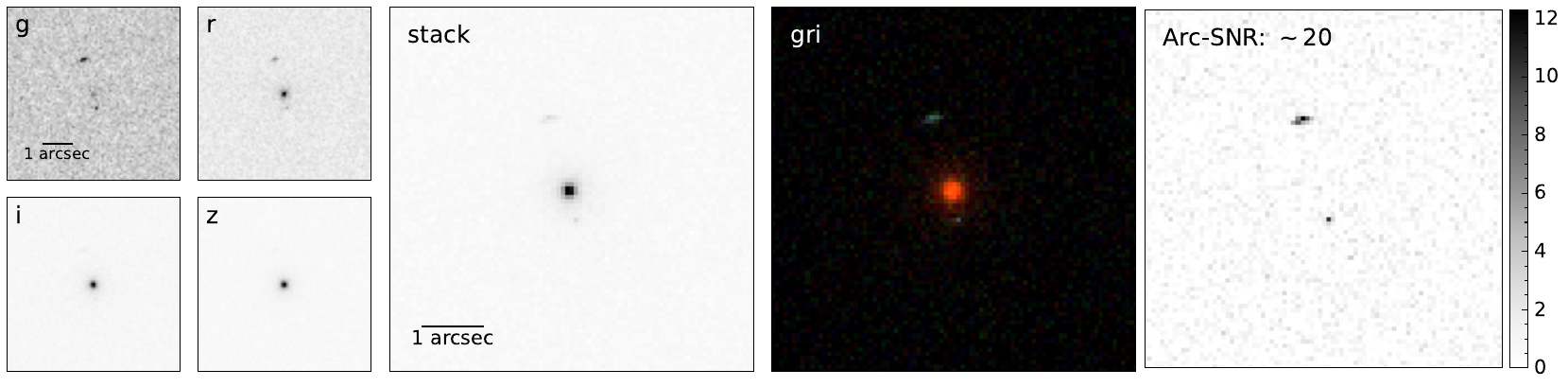}
  \end{subfigure}\hfill
  \begin{subfigure}[t]{\textwidth}
    \centering
    \includegraphics[width=0.9\textwidth]{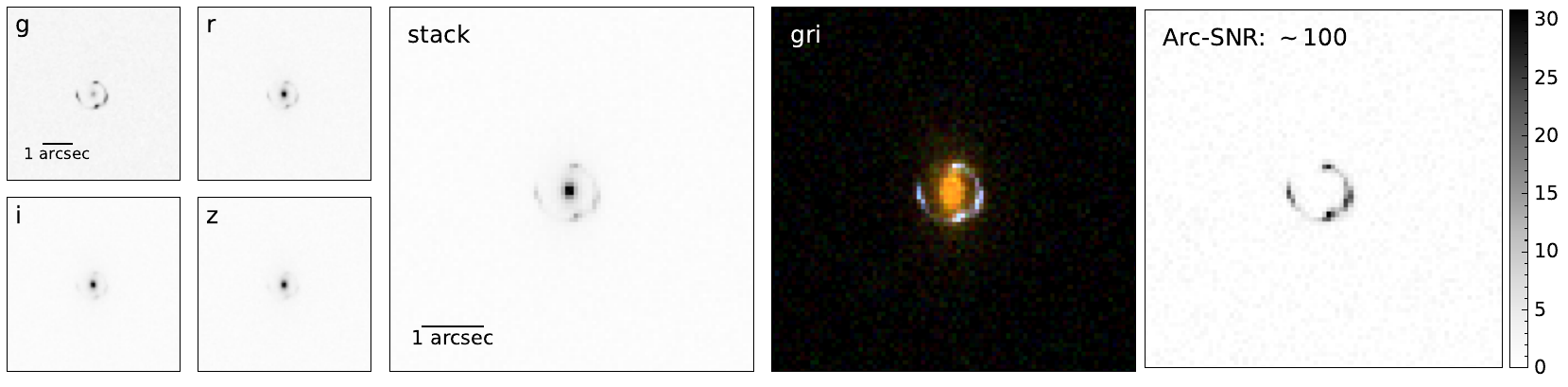}
  \end{subfigure}
  \begin{subfigure}[t]{\textwidth}
    \centering
    \includegraphics[width=0.9\textwidth]{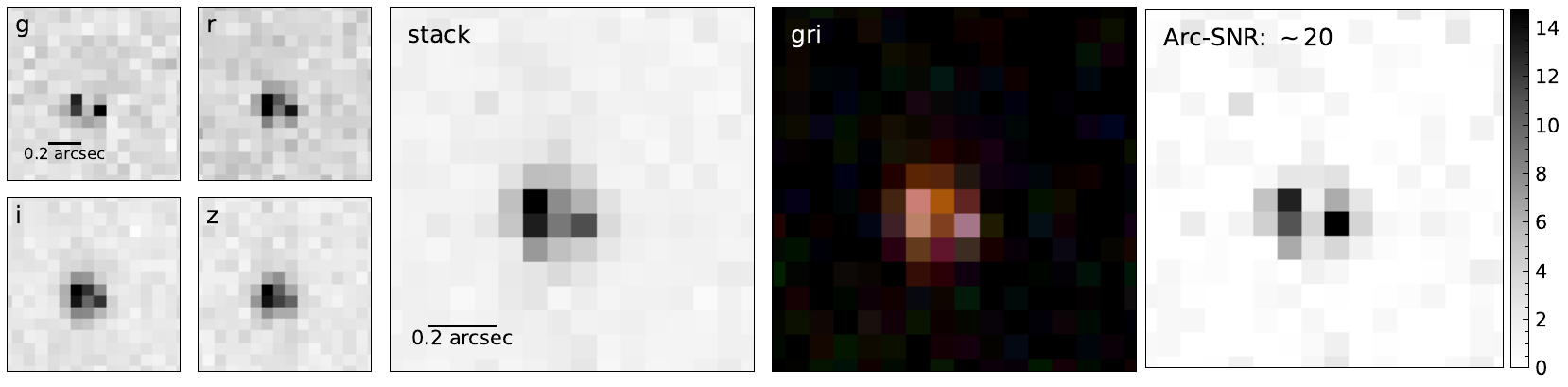}
  \end{subfigure}
\caption {Three lens systems with relatively faint ($\rm SNR_{\rm arc} \approx 20$) lensed images, bright ($\rm SNR_{\rm arc} \approx 100$) lensed images, and small Einstein radius ($\theta_{\rm E} \approx 0.08$) are shown at the top, middle, and bottom rows, respectively. In each row, the left panels show separate images in the $g$, $r$, $i$, and $z$ bands. The middle-left panel displays the stacked image that combines the $g$, $r$, $i$, and $z$ bands. The middle-right panel presents the composite color image. The right panel exhibits the SNR map of the lensed arc for the stacked image.  
}
\label{fig:color_image}
\end{figure*}

\begin{figure*}
     \centering
     \begin{subfigure}[b]{0.3\textwidth}
         \centering
         \includegraphics[width=\textwidth]{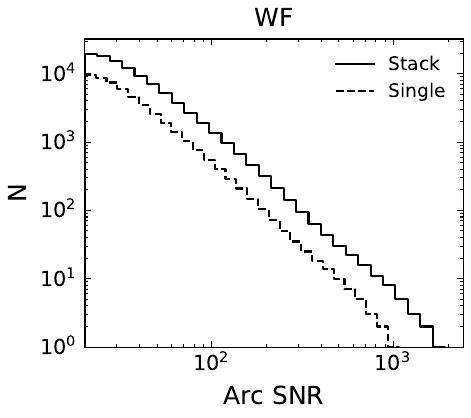}
     \end{subfigure}
     \begin{subfigure}[b]{0.3\textwidth}
         \centering
         \includegraphics[width=\textwidth]{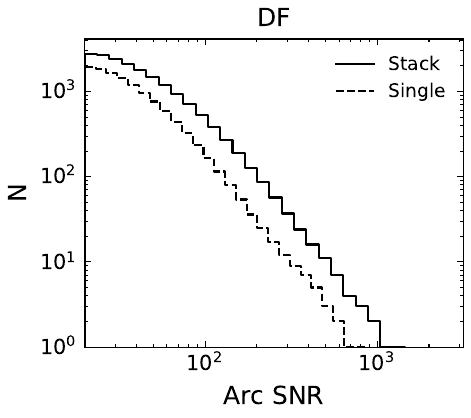}
     \end{subfigure}
     \begin{subfigure}[b]{0.3\textwidth}
         \centering
         \includegraphics[width=\textwidth]{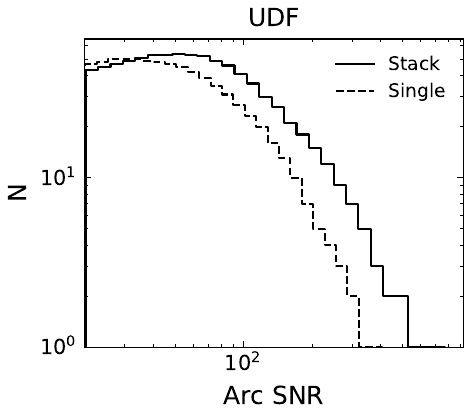}
     \end{subfigure}
\caption {The histograms display the number distribution of the SNR for the lensed arcs in the wide field (left panel), deep field (middle panel), and ultra-deep field (right panel) surveys. The arc SNR measurements are obtained from the stacked image (solid lines) or the single-band image with the highest arc SNR (dashed lines).
}
\label{fig:snr_arc}
\end{figure*}

If the condition for identifying a strong lens is to satisfy the detection requirements outlined in Section~\ref{sec:transfer_func} in at least one image from the $g$, $r$, $i$, or $z$ bands, we can identify approximately 76,000  strong lenses in CSST WF mode. Among these, approximately 41,000 display extended arc structures\footnote{In this work, we say that a strong lens has the so-called ``extended arc structure'' if part of the light within the half-light radius of the source galaxy falls inside the tangential critical line of the lens galaxy.}. It turns out that the lensed features are most prominently detected in the $g$ band, with approximately 78\% of lenses identified through the single-band image exhibiting the highest SNR in the $g$ band. This can be attributed to the fact that high redshift sources often experience star formation and exhibit a bluer color compared to the redder early-type lens galaxies. We refer to this strong lens sample as the ``single-band'' one. Roughly 96\% of the strong lenses in the ideal lens sample did not meet the criteria to be classified as part of the single-band sample due to their failure to satisfy the arc SNR condition outlined in Section~\ref{sec:transfer_func}. However, if we sacrifice the color information and stack the images from the $g$, $r$, $i$, and $z$ bands to enhance the arc SNR, the number of detectable strong lenses will increase to approximately 160,000 with around 82,000 having the extended arc structure. This set of lenses is referred to as the ``stacked'' sample. Our lensing simulation shows that the SNR of the lensed images varies between approximately 20 and 1000 in the stacked sample (refer to the left panel of Figure~\ref{fig:snr_arc}). Among these lenses, approximately 4,000 exhibit bright lensed images with $\rm SNR_{\rm arc}>100$, making them of particular interest as they provide better constraints on the lens's mass (see Section~\ref{sec:discuss}), thus forming the ``golden sample.'' To provide an intuitive demonstration, Figure~\ref{fig:color_image} shows three example lenses from the stacked sample that possess relatively faint lensed images (top row, $\rm SNR_{\rm arc} \approx 20$), bright lensed images (middle row, $\rm SNR_{\rm arc} \approx 100$), and small Einstein radius (bottom row, $\rm \theta_{\rm E} \approx 0.08$), respectively.

\begin{figure*}
  \centering
    \includegraphics[width=0.9\textwidth]{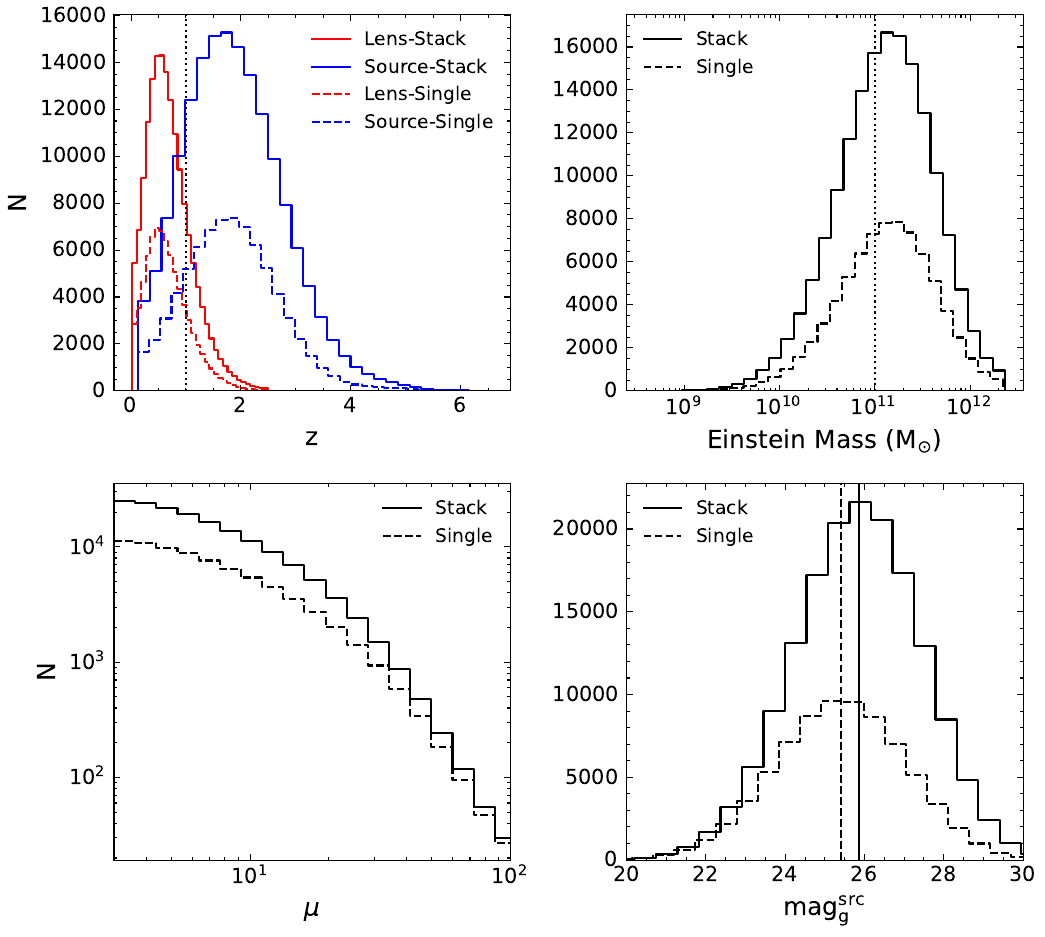}
\caption {
Population statistics of the CSST strong lens samples. \textit{Top left panel}: The red and blue lines show the redshift distribution of the lens galaxies and the source galaxies, respectively.  The solid lines depict the distribution of the lens samples detected via the stacked image, while the dashed lines correspond to those detected using the single-band image with the highest arc SNR. \textit{Top right, bottom left, and bottom right panels}: The histograms show the distribution of the Einstein mass, total magnification, and intrinsic (unlensed) source magnitude in the g-band for both the ``stacked'' (solid lines) and ``single-band'' (dashed lines) samples.
}
\label{fig:csst_lens_properties}
\end{figure*}

Figure~\ref{fig:csst_lens_properties} presents the population properties of the strong lens sample under the CSST WF mode, employing both the ``stacked'' (solid lines) and ``single band'' (dashed lines) detection criteria. The blue and red lines in the top left panel depict the variation of the number of strong lenses with the lens redshift and source redshift, respectively. The vertical dotted line indicates the redshift of 1. In the top right, bottom left, and bottom right panels, histograms show the distribution of the Einstein mass, total magnification, and intrinsic (unlensed) source magnitude. The Einstein mass of deflectors spans from $1.47 \times 10^{9} M_\odot$ to $1.97 \times 10^{12} M_\odot$, with a median value of approximately $1.36 \times 10^{11} M_\odot$. Around $40\%$ of the CSST WF samples have an Einstein mass smaller than $10^{11} M_{\odot}$, as indicated by the vertical dotted line in the top-right panel. This significantly enhances the completeness of the sample for strong lenses in the low-mass range. The total magnification has a median value of $\sim 5$ and exhibits a high-magnification long tail extending to values $\sim 100$. Moreover, the intrinsic (or unlensed) g-band apparent magnitude of the source exhibits median values of approximately 25.40 and 25.87 for the ``single-band'' and ``stacked'' samples, respectively, as illustrated by the vertical lines in the bottom right panel.

\subsubsection{Deep-field lens sample}
\label{sec:DF_mode}
Under the ``single-band'' detection criteria, a search in the CSST DF mode image data yields around $\sim 12,000$ lenses, of which approximately $\sim 5,000$ exhibit extended arc structures. Employing the ``stacked'' detection scheme results in an increase to approximately $\sim 17,800$ and $\sim 7,300$ for these two respective numbers. The middle panel of Figure~\ref{fig:snr_arc} shows the number distribution of lenses with respect to the arc SNR. Other lensing properties (redshift, Einstein radius, magnification, etc) of the lens sample under the CSST DF mode are not significantly different from those of the WF mode. However, the median intrinsic $g$-band source magnitude is approximately $0.98$ magnitude deeper in the DF mode, attributable to the longer exposure time.

\subsubsection{Ultra-Deep-Field lens sample}
\label{sec:UDF_mode}
The CSST UDF survey has the capability to identify $\sim 720$ strong lenses employing the ``single band'' detection approach and $\sim 830$ lenses using the ``stacked'' detection scheme. Within this set, $\sim 280$ lenses showcase an extended arc structure under the ``single band'' criterion, whereas $\sim 320$ lenses exhibit this feature through the ``stacked'' approach. The distribution of arc SNR is illustrated in the right panel of Figure~\ref{fig:snr_arc}, while various lensing properties are shown in Figure~\ref{fig:csst_lens_properties_udf}. Particularly noteworthy, the median $g$-band magnitude of the source in the UDF mode is approximately $0.44$ magnitude deeper compared to the DF mode\footnote{The source catalog used in this work may not be sufficiently deep for the CSST UDF mode.}.

\subsection{Comparing lens samples from different surveys}
\label{sec:res_lens_comparison}
\begin{figure*}
  \centering
    \includegraphics[width=0.9\textwidth]{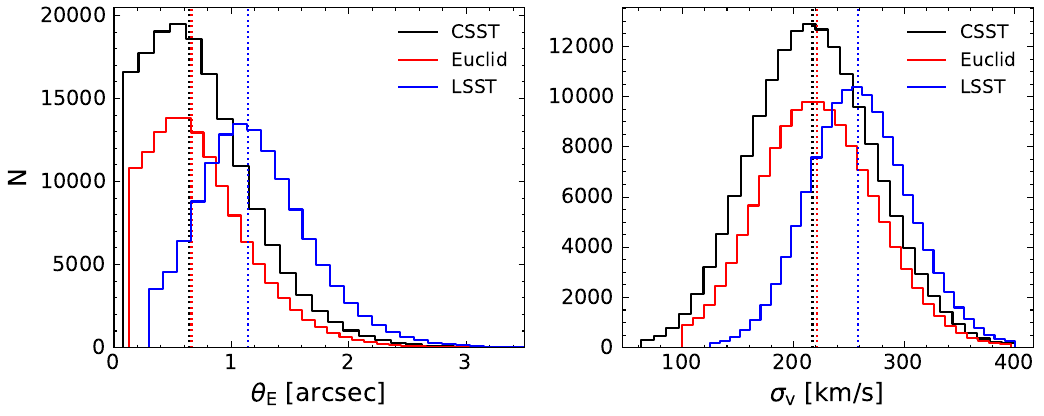}
\caption{The distribution of the lens sample as a function of the Einstein radius (left panel) and the velocity dispersion (right panel) for the CSST (black lines), the LSST (blue lines), and the Euclid (red lines) surveys. Compared to the ground-based LSST, the space-based CSST and Euclid surveys can detect more lenses with smaller Einstein radii and lower velocity dispersions, corresponding to ETGs with lower masses. The vertical dotted lines indicate the median values of each distribution.
}
\label{fig:other_survey}
\end{figure*}

In the upcoming decade, prominent projects such as CSST, Euclid, and LSST will shape the landscape of optical-band image surveys. CSST and Euclid are space telescopes, offering a distinct advantage in spatial resolution compared to their ground-based counterparts like LSST. Conversely, the LSST holds considerable promise for facilitating time-domain studies. 
To enable a comprehensive comparative analysis of lens samples across these three image surveys, we acquire the lensing sample predictions for LSST directly from \cite{Collett15}. Subsequently, we conduct a new lensing simulation tailored to the Euclid survey, employing the latest instrument settings outlined in Table~\ref{table:surveys}. The resulting comparison is visualized in Figure~\ref{fig:other_survey}, illustrating the distribution of lenses relative to both the Einstein radius (left panel) and velocity dispersion (right panel). 
The black histogram represents the results from the CSST WF mode using the ``stacked'' detection scheme. The red histogram represents the one from the Euclid survey by searching strong lenses in the $VIS$ band image. The blue histogram shows the lens sample distribution from the LSST survey, using the ``optimal stacked'' scheme\footnote{The seeing of a ground-based telescope varies with time, and images from multiple exposures of the same target may be blurred differently. For lens searching, it is possible to select only all individual exposures that are not badly blurred for stacking to identify the lensed arc better. This choice trades the image SNR for spatial resolution, which is called the ``optimal stack'' scheme in \cite{Collett15}.} for lens searching \cite{Collett15}. 
CSST, Euclid, and LSST are forecasted to detect approximately \num{160000}, \num{132000}, and \num{129000} lenses, respectively. The velocity dispersion of the CSST lens sample is $217.19\pm50.55 \, \text{km/s}$ (median$\pm$standard deviation) similar to that from Euclid's result, $221.10 \pm 48.18 \, \text{km/s}$. The lens sample of CSST exhibits Einstein radii of $0.64\pm0.42 \arcsec$, slightly smaller than Euclid's $0.66\pm0.40 \arcsec$. However, these values remain comparable, suggesting the two surveys have similar lens populations. LSST lenses show a velocity dispersion of $258.09\pm41.62 \, {\rm km/s}$ and an Einstein radius of $1.14\pm0.42 \arcsec$; thus, its selection function favors more massive lenses.

%% file: discuss.tex
\section{Discussion}
\label{sec:discuss}
We have demonstrated that the CSST can potentially discover approximately 160,000 GGSLs, although this number may vary with different lens detection criteria, as discussed in Section~\ref{sec:discuss_1}. Section~\ref{sec:discuss_2} assesses the measurement accuracy of the lensing galaxy's mass using the CSST data. Understanding the population differences between the parent and lensed samples (i.e., the selection effect) is crucial for many lensing applications, which is addressed in Section~\ref{sec:discuss_3}. A notable feature of the CSST survey is its ability to identify many exotic lens systems that are rare in current samples, as highlighted in Section~\ref{sec:discuss_4}. Furthermore, the CSST and Euclid imaging data are complementary regarding filter coverage; Section~\ref{sec:discuss_5} explores how to coordinate CSST and Euclid strong lens data for improved strong lens science.

\subsection{Assessing Uncertainties in the Number of Predicted Strong Lenses}
\label{sec:discuss_1}
The major source of uncertainties that could significantly change the number of GGSLs predicted by our simulation is the choice of lens finder, as demonstrated by \citet{Collett15}. In this work, we have deliberately adopted the same detectability condition as \citet{Collett15} (thus the same lens finder) for a consistent comparison. These detectability conditions are proved to be reasonably good by visually inspecting the lensing image. Nevertheless, we still try three different choices to tighten the detection criteria for a more thorough investigation. We raise the SNR threshold from 20 to 40, increase the magnification threshold from 3 to 4, and modify the image-resolved threshold from $\theta_{\mathrm{E}}^2 > (R_{\rm e}^{s})^2 + (s / 2)^2$ to $\theta_{\mathrm{E}}^2 > (R_{\rm e}^{s})^2 + s^2$. We find the number of detectable lens drops by 77\%, 26\%, and 2\%, respectively, While the change in arc SNR is the primary factor affecting the number of detectable lenses. Visually inspecting the lens-light subtracted image shows that the lensed features are already evident when the total arc SNR reaches 20.

It is worth noting that the lens finders we have employed thus far rely on either single-band images or stacked ones, failing to fully exploit the color information provided by the CSST. Since lenses and sources usually have distinct colors, leveraging this feature could facilitate the deblending of lens and source signals. Consequently, it could enable the detection of more lenses, particularly for those with small Einstein radii. To assess the impact of color information, we adopt the methodology proposed by \cite{Gavazzi2014}, which uses color difference imaging between the $g$ and $i$ bands for lens identification. In essence, this method rescales the $i$ band image by a factor of $\alpha$ to align its lens light with that of the $g$ band. Since the source tends to be bluer than the lens, the lensed arc manifests as a positive residual in the $g - \alpha i$ image, while the lens light is effectively subtracted. Applying the detection criterion outlined in Section~\ref{sec:transfer_func} to the $g - \alpha i$ image reveals the detection of $\sim 54,000$ strong lenses. Interestingly, all of these $\sim 54,000$ lenses are already part of the single-band or stacked sample—indicating that color information does not significantly aid in lens finding. This phenomenon arises from our idealized assumption regarding the lens finder: perfect subtraction of lens light down to the noise level. In practice, achieving this level of lens light subtraction remains challenging due to the complex morphology and high blending of the lens light and source light. Although a neural network using all multi-band images from CSST could offer a more realistic approach to lens finding, the associated tasks of training, tuning, and understanding the selection effects of such a network require substantial effort beyond the scope of this paper (such as \citet{Herle2023, Leuzzi2024}). We shall leave it to future works.

\subsection{Probing Lens Galaxy Masses with CSST Strong Lensing Data}
\label{sec:discuss_2}
\begin{figure}
  \centering
    \includegraphics[width=\columnwidth]{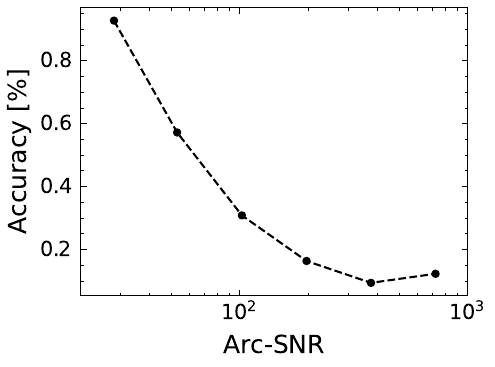}
\caption {Lens modeling accuracy for the Einstein radius is explored in relation to the total SNR of the lensed arc. The SNR range, from 20 to 1000, is divided into six equally spaced logarithmic bins, each containing 30 lenses with diverse lensing configurations. Black dots indicate the median modeling accuracy for each SNR bin. This accuracy accounts for both biases and statistical uncertainties. Notably, the modeling accuracy in the highest SNR bin shows a slight degradation, attributed to the more pronounced adverse impact of the imperfect PSF for the stacked image.}
\label{fig:model_accuracy}
\end{figure}

For gravitational lenses identified through the CSST wide-field survey using stacked images, the SNR of the lensed arcs varies from $\sim20$ to $\sim1000$. To assess the precision with which the lensed arc can constrain the lens mass model, we model the lensed arc images generated by our lensing simulation with an SIE lens plus Sersic source model\footnote{Images of lens modeling results for several example lenses are given in Appendix~\ref{sec:appx_c}.}, assuming the lens light has been subtracted to the noise level. We find the Einstein radius can be pinned down to $\sim 1\%$ and $\sim 0.1\%$ levels (see Figure~\ref{fig:model_accuracy}), depending on the SNR of the lensed arc. There is a counter-intuitive trend where the modeling accuracy deteriorates for the highest SNR bin. We attribute this to the imperfect PSF in the stacked image, as the PSF size is oversimplified by taking the mean FWHM across the $g$, $r$, $i$, and $z$ bands. Future analyses can circumvent this issue by directly performing multi-band lens modeling instead of working on the stacked image. This will mitigate the impact of an imperfect PSF while preserving the SNR enhancement provided by multi-band data. It is important to note that the reported precisions here do not account for various systematic effects that can affect the accuracy of lens modeling, such as missing complexities in the mass model \citep{mass_complex_1,mass_complex_2,mass_complex_3,mass_complex_4,mass_complex_5,mass_complex_6,mass_complex_7}, inaccurate PSF \citep{psf_error}, imperfect subtraction of the lens light \citep{Nightingale2022}, and correlated CCD noise when working with real lens data. Nevertheless, these values provide insights into the statistical ability of lensed arcs to constrain the mass of the lens.

\subsection{Characterizing the Selection Effects in CSST Strong Lensing Observations}
\label{sec:discuss_3}
\begin{figure*}
  \centering
    \includegraphics[width=\textwidth]{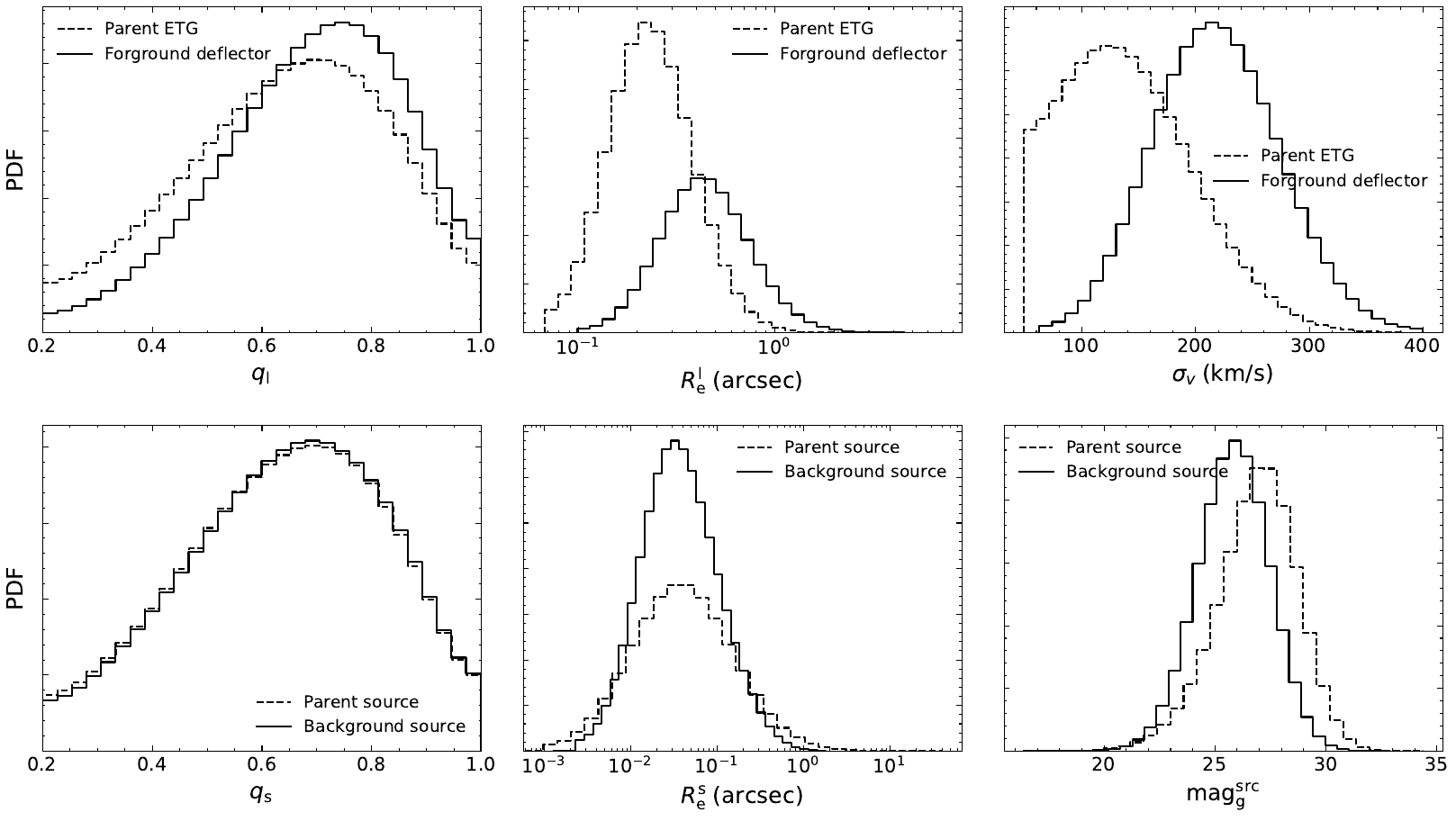}
\caption {This figure illustrates the lensing selection effect through a comparative analysis of the properties of the lens sample and its parent population. The top panels, from left to right, display solid lines representing the probability density functions (PDFs) of the lens galaxy's axis ratio, effective radius, and velocity dispersion, while the corresponding PDFs of the parent population are shown as dashed lines. In the bottom panels, from left to right, we compare the differences in the PDFs between the lens and parent populations for the source's axis ratio, effective radius, and g-band magnitude.}
\label{fig:selection_effect}
\end{figure*}

Understanding the selection function in CSST GGSL observations is crucial for making reliable statistical inferences from a large number of lenses \citep{Sonnenfeld22c, Zhou2024}. Figure~\ref{fig:selection_effect} presents the probability density distribution of various lensing properties for both lensed (solid lines) and parent population (dashed lines). The top panels reveal that ETGs acting as lenses are typically more massive (as evidenced by their higher velocity dispersions) and have larger sizes (as manifested by their effective radii). Additionally, the lensing ETGs tend to be rounder in shape, consistent with findings from other lensing studies. The lower panels highlight that lensed sources exhibit higher luminosity than the parent sample, attributable to the depth limitations of CSST. Unlike the lens galaxies, the source exhibits no substantial axis-ratio selection effect. The size distribution of lensed sources peaks around $\sim 0.09 \arcsec$, with fewer instances of exceptionally small or large sources than in the parent sample. This outcome emerges from the interplay of two factors. When the source size is exceptionally small, it typically appears faint due to the applied scaling relation (see equation~\ref{eq:background_Re}), resulting in a lensed arc lacking the required SNR for successful detection. Conversely, if the source size is markedly large, it may fail to meet the ``resolved condition'' criteria outlined in Section~\ref{sec:transfer_func}, preventing the identification of counter images produced by strong lensing.

\subsection{Uncovering Rare Strong Lenses in CSST}
\label{sec:discuss_4}
Among those $\sim$160,000 CSST lenses, some systems that are either rare or undetected in previous lens samples, such as high-redshift ($z_l > 1$) or low-mass ($M_{ein} < 10^{11} M_{\odot}$), draw particular interest since they represent ``place in parameter space'' that are uncovered by the current sample. Our simulations predict the CSST could find $\sim$\num{31000} high-redshift and $\sim$\num{62000} low-mass lenses, significantly increasing the completeness of the GGSL sample.

Furthermore, CSST has the potential to detect unusual systems such as spiral galaxy lenses and Double Source-Plane Lenses (DSPLs). Regarding spiral galaxy lenses, estimating them accurately is a complex task that requires the development of a dedicated population model specific to spiral galaxies. Although constructing such a model exceeds the scope of this study, it is possible to provide a rough estimate of the number of spiral galaxy lenses by considering the proportion of existing spiral galaxy lenses. For instance, the SLACS project identified approximately $\sim 10\%$ (10 out of 85) of GGSLs as spiral galaxy lenses. By extrapolating this ratio, it is plausible to suggest that the CSST could detect around $\sim 15,000$ spiral galaxy lenses. In the context of DSPLs, where two sources at different redshifts produce distinct images, extending our simulation code to incorporate multiple sources is straightforward. Specifically, we randomly select 50 sources from the mock catalog for each putative lens galaxy and distribute them around the lens. The spatial density of these 50 sources aligns with that of the cosmological simulation. Following the criteria outlined in Section~\ref{sec:sim_ideal_lens}, we register systems in which two background sources are strongly lensed\footnote{We disregard the lensing effect caused by the foreground source that is closer to the observer.}. This process yields 172,500 physical DSPLs over a sky area of 17,500 square degrees. Next, we focus on systems where both sets of lensed images formed by two sources are observable. Based on the criteria described in Section~\ref{sec:transfer_func}, we predict that the CSST wide-field survey can detect $\sim 250$ DSPLs. Combined with our prediction of $\sim$160,000 observable single source-plane lenses, this suggests an average occurrence of one DSPL in $\sim 640$ strong lenses for the CSST wide-field survey\footnote{\citet{euclid_double_ring} predicts that the Euclid survey could discover around 1,700 DSPLs, an order magnitude higher than our prediction. This substantial difference is surprising, especially considering the similar depth and resolution between the CSST and Euclid, as well as the comparable number of predicted single-source lenses. We are currently unable to address the reason for this discrepancy. However, we present our number prediction for DSPLs as reasonable, as their strong lensing rates are consistent with the single-source simulation result. See Appendix \ref{sec:appx_e} for more details.}.

\subsection{Synergizing CSST and Euclid Strong Lensing Observations}
\label{sec:discuss_5}
Our simulations reveal that the population statistics of CSST and Euclid strong lenses are similar, which is expected given their comparable spatial resolution and depth. However, the smaller PSF and pixel sizes of CSST provide certain advantages in detecting small-mass lenses. Since both telescopes cover nearly identical sky regions, effective coordination in utilizing the strong lensing data from these two surveys is crucial. In the future, CSST and Euclid are expected to discover hundreds of thousands of strong lenses, making the acquisition of spectroscopic redshifts a costly endeavor. Unlike Euclid, CSST possesses optical color information ($ugriz$), enabling the calculation of photometric redshifts for lens and source galaxies. However, CSST's lack of an infrared band can significantly impact the measurement accuracy of photometric redshifts, particularly for high-redshift sources ($z>1$); Euclid's near-infrared (NIR) band can help mitigate this limitation. A joint analysis using CSST and Euclid's image data can be conducted, akin to stacking images from two separate surveys to enhance the SNR and yield more robust constraints on lens models.

\begin{table}
\centering
\caption{We present the number of detectable lenses using various detection schemes. These schemes utilize single-band images (referred to as ``single''), stacked images (referred to as ``stack''), or a combination of both (referred to as ``combine''). They are categorized under various CSST survey modes (WF, DF, and UDF). Additionally, we provide the number of lenses that exhibit bright lensed images (with $\rm SNR_{\rm arc}>100$), have a lens redshift exceeding 1, and possess an Einstein mass lower than $10^{11} M_\odot$. The predicted numbers shown here are subject to uncertainty arising from Poisson counting noise, meaning $\sim 10,000$ lenses have an associated uncertainty of $\sim 100$ lenses.}
\label{table:numbers}
\begin{tabular}{cccc}
\hline
                           & WF     & DF    & UDF  \\
\hline
stack                      & 161000  & 17800  & 830  \\
single                     & 76300   & 11900  & 720 \\
combine                    & 162000  & 18100  & 840 \\
$\rm SNR_{\rm arc}>100$    & 4000    & 1010   & 200 \\
$z_l > 1$                  & 30600   & 4200   & 200  \\
$M_{ein} < 10^{11}M_\odot$ & 61500   & 6900   & 310 \\
\hline
\end{tabular}
\end{table}

%% file: summary.tex
\section{Summary}
\label{sec:summary}
In this paper, we performed a Monte Carlo simulation that adopts a realistic population for the lens and source properties to predict how many GGSLs would be observed by CSST and explore the population statistics of those CSST lenses. Our simulations consider various selection functions, including the telescope PSF, the survey depth, and the lensing magnification bias \citep{kochanek2004saas}.

The CSST can potentially discover approximately 160,000 GGSLs through the wide-field survey using the ``stacked image'', assuming a Poisson-limited lens light subtraction. The number of lenses was reduced to approximately 17,900 and 840 for the CSST deep-field and ultra-deep-field surveys, respectively. By imposing the condition that the lensing feature must be detected in at least one of the individual $g$, $r$, $i$, and $z$ bands, the CSST wide-field, deep-field, and ultra-deep-field surveys would detect approximately 76,400, 12,000, and 730 GGSLs, respectively. To facilitate the reader's accessibility, the information on the number of detectable lenses for CSST is compiled in Table~\ref{table:numbers}.

The spatial resolution and SNR attainable in future CSST lensing images enable inference of the lens mass at the percent level, assuming other potential systematics have been addressed. Performing a joint analysis of CSST and Euclid's image data—simultaneously modeling both datasets—can further enhance the accuracy of the lens model. Additionally, measurements of photometric redshifts and stellar population synthesis models can be improved by using the compensated photometry information of CSST's optical band and Euclid's near-infrared bands. Therefore, coordinated efforts between CSST and Euclid in handling strong lensing data are highly appreciated.

%% file: appx_A.tex
\section{The FWHM of CSST's PSF}
\label{sec:appx_a}
\begin{figure*}
  \centering
    \includegraphics[width=\textwidth]{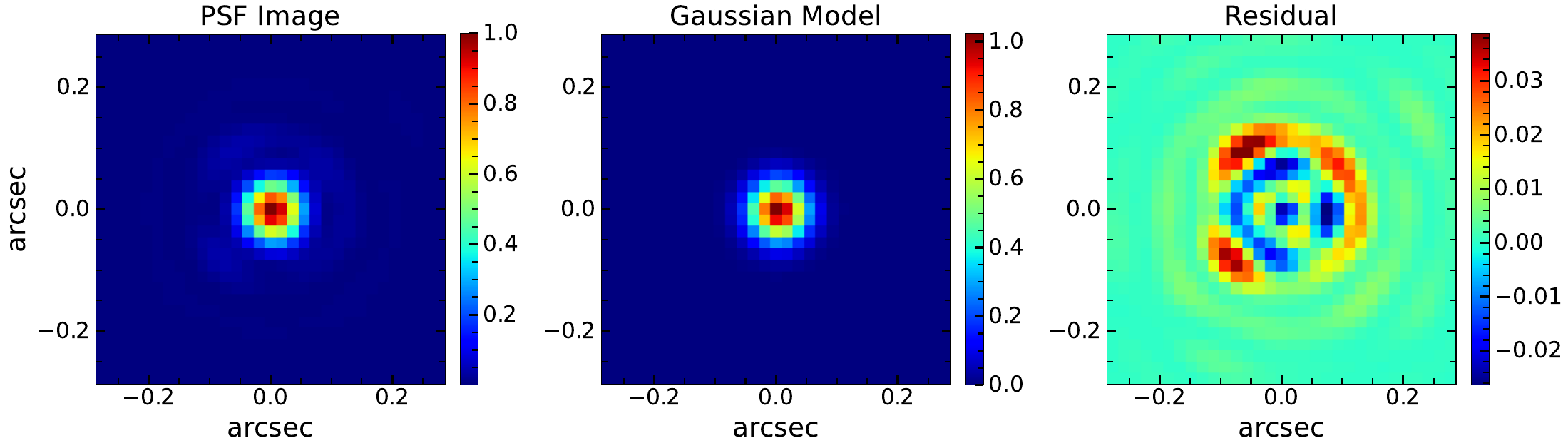}
\caption {To obtain the FWHM value, a Gaussian model is employed to fit the simulated PSF image of CSST. Panels from left to right display the simulated PSF image, the image obtained through Gaussian model fitting, and the residual image calculated as the difference between the left and middle panels. Images shown here have a 4 by 4 supersampling, resulting in a pixel size of $0.074/4=0.0185 \arcsec$.}
\label{fig:csst_psf_fit}
\end{figure*}

\begin{figure}
  \centering
    \includegraphics[width=\columnwidth]{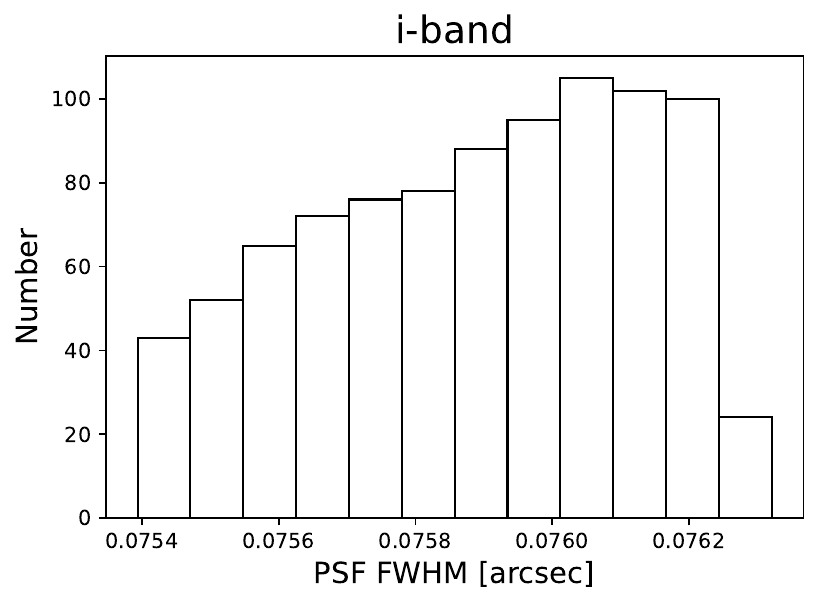}
\caption {The distribution of FWHM values for the PSF of CSST in the $i$ band. The variation in PSF size is caused by changes in position within the CCD focal plane.}
\label{fig:i_fwhm_hist}
\end{figure}

The CSST uses the $R_\mathrm{EE80}$ as the primary metric to define the PSF properties, whereas other telescopes like the Hubble Space Telescope (HST) and Euclid typically employ the FWHM. To enable a direct comparison between different telescopes, we determine the FWHM by fitting a Gaussian model to a set of simulated PSF images captured by the CSST. Figure~\ref{fig:csst_psf_fit} shows an example of PSF fitting in the $i$ band. Figure~\ref{fig:i_fwhm_hist} presents the distribution of PSF FWHM in the $i$ band, illustrating the slight variation in PSF size across the CCD focal plane. The reported FWHM value of the PSF in Table~\ref{table:surveys} corresponds to the median.

For a comprehensive crosscheck, we also attempted to apply Gaussian model fitting to an HST PSF in the F814W filters. The resulting FWHM value was approximately $0.09 \arcsec$, which closely resembles that of the CSST $z$ band ($\sim 0.089 \arcsec$). This result aligns with our expectations, as both the HST and CSST are two-meter telescopes, and the F814W filters closely correspond to the CSST $z$ band in terms of wavelength.

%% file: appx_B.tex
\section{Other lensing properties}
\label{sec:appx_b}
\begin{figure}
  \centering
    \includegraphics[width=\columnwidth]{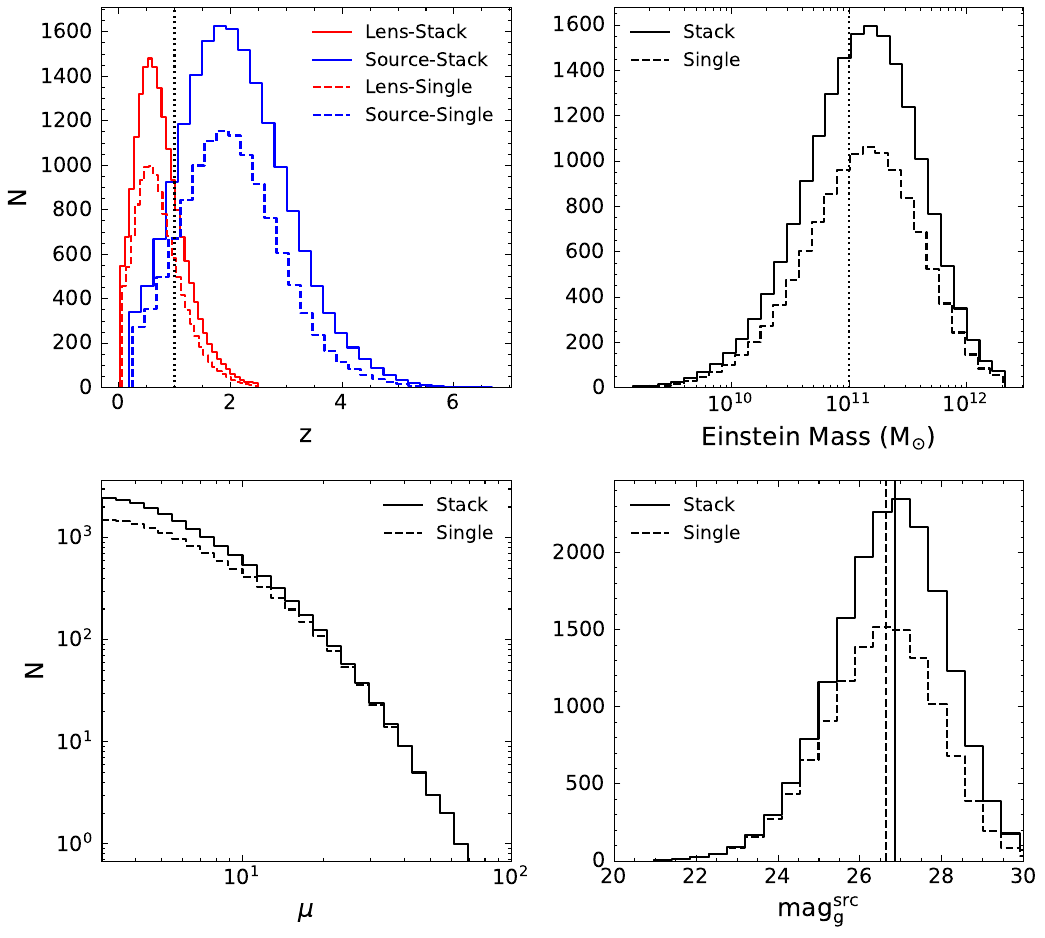}
\caption {Similar to Figure~\ref{fig:csst_lens_properties}, but for the CSST deep field surveys.}
\label{fig:csst_lens_properties_df}
\end{figure}

\begin{figure}
  \centering
    \includegraphics[width=\columnwidth]{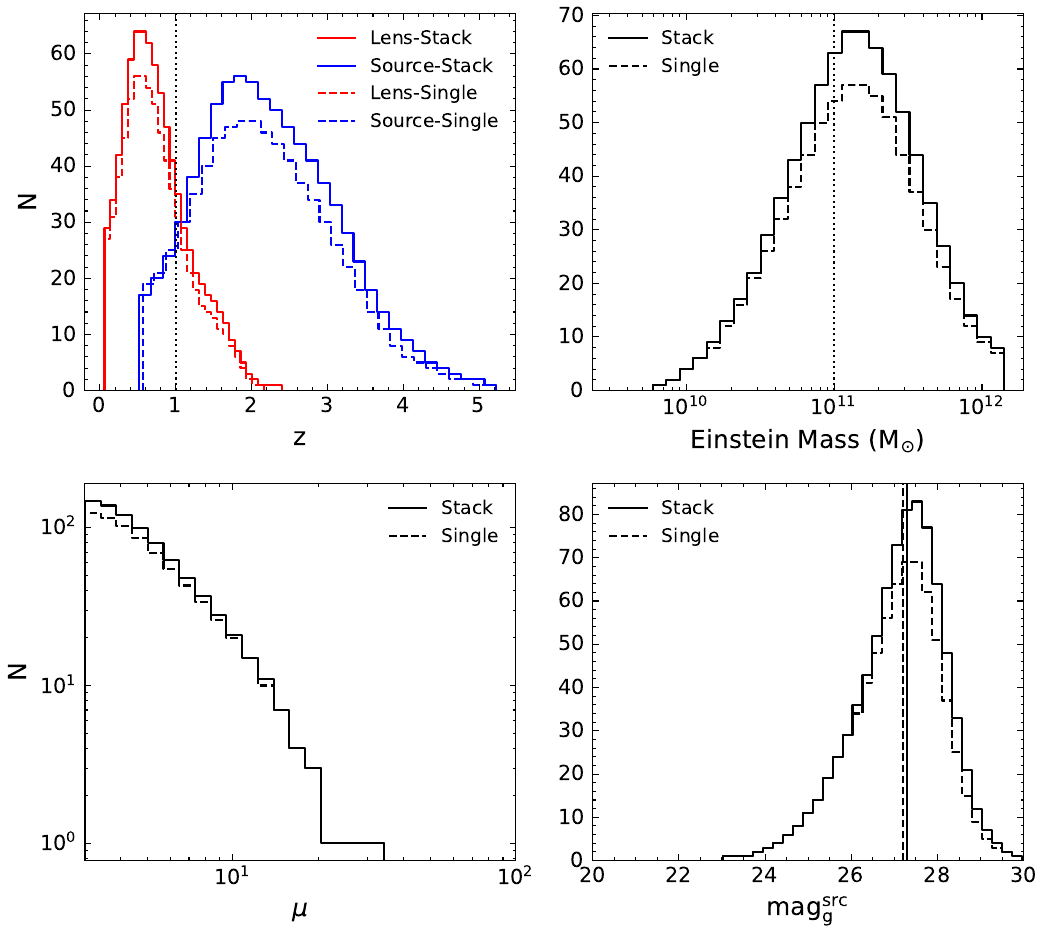}
\caption {Similar to Figure~\ref{fig:csst_lens_properties}, but for the CSST ultra-deep field surveys.}
\label{fig:csst_lens_properties_udf}
\end{figure}

The population statistics of the lens sample for the CSST deep field and ultra-deep field surveys are presented in Figures~\ref{fig:csst_lens_properties_df} and ~\ref{fig:csst_lens_properties_udf}, respectively. 

%% file: appx_C.tex
\section{Remark on the depth of CSST and Euclid}
\label{sec:appx_c}
The CSST boasts a telescope diameter of two meters, giving it a significant advantage in terms of depth when compared to Euclid, which has a smaller one-meter diameter. However, it's worth noting that Euclid compensates for its smaller size with longer exposure times of $565\times3$s for its wide field survey, whereas CSST employs shorter exposure times of $150\times2$s. Moreover, the $VIS$ imager on Euclid is equipped with a broad-band filter that effectively covers the $riz$ bands present in CSST. As a result, making an apple-to-apple comparison between the two telescopes becomes somewhat indirect due to these contrasting factors. We shall consider two examples here to illustrate the depth of CSST and Euclid, in particular for our lensing simulation. 

To begin, let's consider a source exhibiting an apparent magnitude of 25 in the $riz$ bands. According to the approximation outlined in \cite{Collett15}, the magnitude in the $VIS$ band is also 25, as it represents the mean of the $riz$ bands. Consequently, CSST's 300s exposures will generate \num{1823} electron count on the CCD, whereas Euclid, with a longer exposure time of 1695s, will generate \num{2092} electron counts. With a higher number of electron counts, Euclid shows a larger signal-to-noise ratio, taking into account the Poisson noise. Thus, in terms of the $VIS$ band, Euclid's wide-field survey offers slightly greater depth compared to CSST's.

\begin{figure*}
  \centering
    \includegraphics[width=\textwidth]{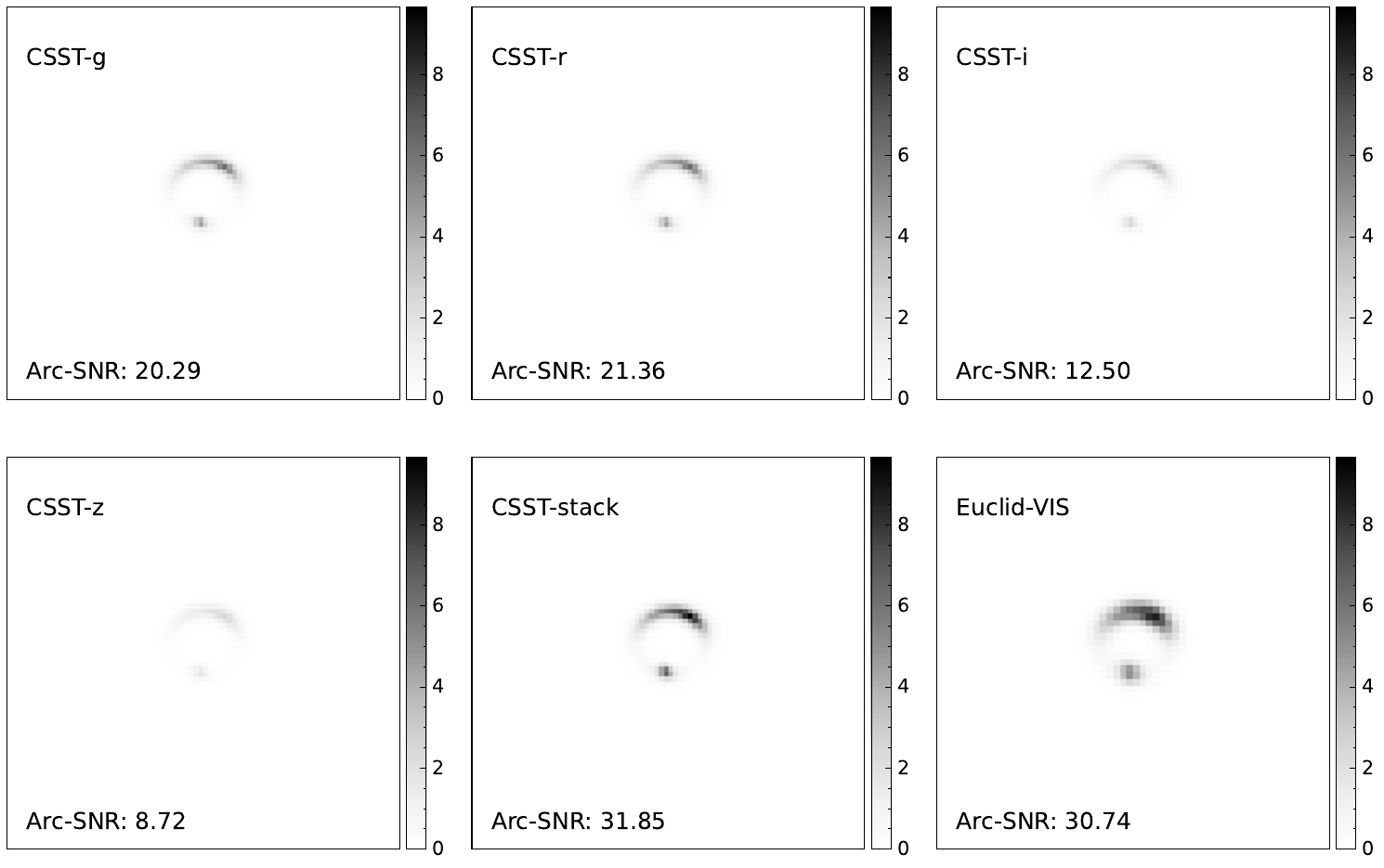}
\caption {The ideal signal-to-noise ratio (SNR) map of the lensed arc is generated without noise realization. It is computed for the CSST filters $g$, $r$, $i$, $z$, the stack image of $griz$, and the Euclid $VIS$ image. The total SNR of the lensed arc is provided in the legend for each filter. 
}
\label{fig:csst_euclid_imaging_depth}
\end{figure*}

Next, let us explore the appearance of observable lenses in our CSST lensing simulation within the context of the Euclid survey. Figure~\ref{fig:csst_euclid_imaging_depth} presents a typical example illustrating this comparison. The panels, arranged from top-left to bottom-right, depict the ideal SNR maps of the lensed arc (representing the noise map prior to incorporating a noise realization) for CSST-$g$, CSST-$r$, CSST-$i$, CCSST-$z$, CSST-$stack$, and Euclid $VIS$ bands, respectively. The legend provides the total SNR values corresponding to the lensed arc. Remarkably, our analysis indicates that CSST and Euclid yield similar depths in terms of the detected total SNR of the lensed arc. However, CSST exhibits a slightly higher SNR value, indicating a better depth of observation. This advantage arises from the stacking of images from the $g$, $r$, $i$, and $z$ bands, as well as the smaller PSF of CSST that results in sharper images.

It is important to note that our CSST lensing simulation solely incorporates the $griz$ bands. This limitation arises from the fact that the source catalog we utilized currently lacks data from other bands. However, it is worth considering that by stacking the image data from the $u$ and $NUV$ bands, we may potentially enhance the number of detectable lenses within CSST.

%% file: appx_D.tex
\section{Lens modeling results}
\label{sec:appx_d}
\begin{figure*}
  \centering
    \includegraphics[width=\textwidth]{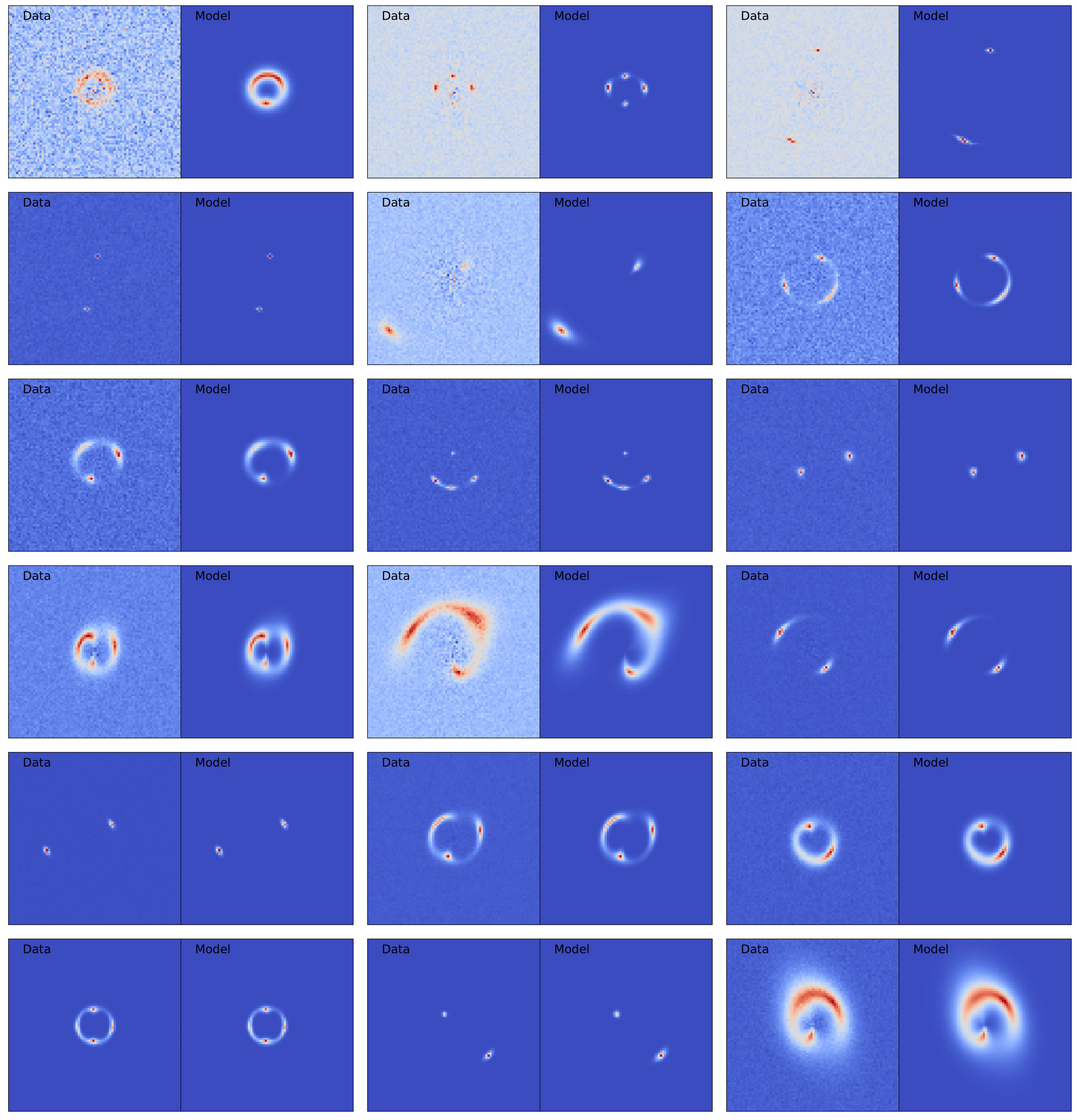}
\caption {From top to bottom rows, the SNR of the lensed images is at the level of approximately 28,  53,  102, 196,  376, and  722, corresponding to the six equally spaced logarithmic bins employed in Figure~\ref{fig:model_accuracy}. Each row contains three panels displaying the lens modeling results of three lenses with distinct lensing geometries. The data and model images are shown on the left and right, respectively.}
\label{fig:csst_mock_modeling}
\end{figure*}

The lensed images generated by our lensing simulation, after subtracting the lens light, are modeled using an SIE lens plus a Sersic source model. The modeling results for several example lenses are arranged in rows from top to bottom in Figure~\ref{fig:csst_mock_modeling}, based on the SNR of the lensed images. Each row consists of three panels displaying the modeling results of three example lenses. These lenses have similar SNR levels but different lensing geometries, with the data and model images shown on the left and right, respectively.

%% file: appx_E.tex
\section{Double Source-Plane Lensing Rate}
\label{sec:appx_e}
For a given ETG with physical properties represented by a vector $\vec{\eta}$, the probability that it strongly lenses a single background source and produces a strong lensing event observable by CSST is denoted as $P(\vec{\eta})$. Since sources are randomly distributed across the sky and can be considered independent of each other, the probability of producing a CSST-observable DSPL event for a given ETG is simply $P(\vec{\eta})^2$. In principle, all the physical information encapsulated in $\vec{\eta}$—such as the brightness distribution of the ETG and the ellipticity of the ETG's mass—can impact the lensing rate $P(\vec{\eta})$. Among these factors, the most crucial one is the velocity dispersion of the lens ($\sigma_v$), which essentially determines the strong lensing cross-section \citep{lensing_textbook}. To simplify the problem, we assume that velocity dispersion solely determines the lensing rate, so $P(\vec{\eta}) \approx P(\sigma_v)$. Under this assumption, the lensing rate for DSPLs is given by $P(\sigma_v)^2$.

\begin{figure*}
  \centering
    \includegraphics[width=\textwidth]{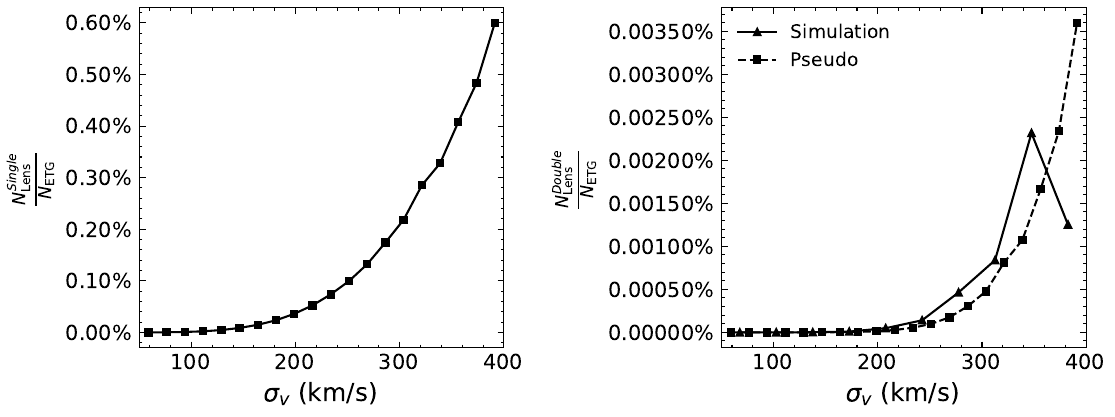}
\caption{
The lensing rate, defined as the ratio of the number of observable lenses to the parent ETGs used to simulate them, is plotted as a function of velocity dispersion. Solid lines in the left and right panels represent results from our Monte Carlo simulations for single-source and double-source strong lensing, respectively. The dashed line in the right panel corresponds to a `pseudo' double-source lensing rate obtained by squaring the curve from the left panel. The solid line in the right panel exhibits a counter-trend drop in the largest velocity dispersion bin due to significant Poisson counting noise.
}
\label{fig:DSPL_rate}
\end{figure*}

In the left panel of Figure~\ref{fig:DSPL_rate}, we present the lensing rate $P(\sigma_v)$ as a function of velocity dispersion for single-source lenses. $P(\sigma_v)$ can be straightforwardly calculated from our Monte Carlo simulation results: $P(\sigma_v) = \frac{N_{\text{lens}}^{\text{Single}}}{N_{\text{ETG}}}$, where $N_{\text{lens}}^{\text{Single}}$ and $N_{\text{ETG}}$ represent the number of observable lenses and ETGs in each velocity dispersion bin, respectively. Similarly, the solid line in the right panel shows the lensing rate of DSPLs as a function of velocity dispersion, calculated from our double-source lensing simulation. The dashed line represents the `pseudo' double-source lensing rate, obtained by squaring the curve from the left panel. The consistency between the dashed and solid lines indicates that the lensing rate from our single-source and double-source simulations aligns well. Therefore, our number prediction for DSPLs is reasonable in terms of the lensing rate.

